\documentclass{article}

\usepackage{amsmath}
\usepackage{amsfonts}
\usepackage{amssymb}
\usepackage{caption}
\usepackage{subcaption}
\usepackage{graphicx}
\usepackage[top=2.5cm, left=2.5cm, bottom=2.5cm, right=2.5cm]{geometry}

\begin{document}

\title{Optical realization of nonlinear quantum dynamics}

\author{F. Soto-Eguibar,$^{1,*}$ V. Arrizon,$^1$ A. Z\'u\~niga-Segundo,$^{2}$ and H.M. Moya-Cessa$^{1}$\\
$^1${\small Instituto Nacional de Astrof\'{\i}sica, \'Optica y Electr\'onica.} \\
\small {Calle Luis Enrique Erro No. 1, Santa Mar\'{\i}a Tonantzintla, Puebla, 72840 Mexico}\\
$^2${\small Departamento de F\'{\i}sica, Escuela Superior de F\'{\i}sica y Matem\'aticas, IPN},\\
 {\small Edificio 9, Unidad Profesional Adolfo L\'opez Mateos, DF, 07738 Mexico} \\
$^*$\small {Corresponding author: feguibar@inaoep.mx}}

\maketitle

\begin{abstract}
In a cavity filled with a Kerr medium it is possible to generate the superposition of coherent states, i.e. Schr\"{o}dinger cat states may be realized in this system. We show that such a medium may be mimicked by the propagation of a conveniently shaped Gaussian beam in a  GRIN device. This is attained by introducing a second order correction to the paraxial propagation of the beam. An additional result is that a Gaussian beam propagating in GRIN media, may split into two Gaussian beams.
\end{abstract}

\textbf{Keywords:} electromagnetic wave propagation, quadratic GRIN media, quantum Kerr media, Classical-Quantum analogies.\\

Conventional applications of a GRIN medium are focusing and image formation \cite{gomez}. Additionally, it has been established that a GRIN medium can support invariant propagation modes, either in the paraxial \cite{ozaktas} and the non paraxial domains \cite{ojeda}.  In a different context, light propagation in a GRIN medium can be employed as a form of optical emulation of certain quantum phenomena. An example is the mimicking of quantum mechanical invariants by the propagation of light through the interface of two coupled GRIN devices \cite{Arr}. Cross applications between quantum mechanics and classical optics are common due to the fact that the Schr\"{o}dinger equation and the paraxial wave equation in classical optics are formally equivalent. One can extend the application in order to consider not only the paraxial regime but also the non-paraxial one, i.e. the complete Helmholtz equation. For instance, supersymmetric methods, common to quantum mechanics, have been proposed in classical optics \cite{Chuma,Dimi,David}. 

The generation of nonclassical  states of the electromagnetic field is a subject of much interest in quantum optics, not only because of their fundamental implications, but also because of possibilities of application that they convey. One may mention among the several nonclassical states found in the literature: (a) squeezed states \cite{loudon},  (b) the particularly important limit of extreme squeezing; i.e., Fock or number states \cite{4}, (c) macroscopic quantum superpositions of quasiclassical coherent states \cite{1} with different mean phases or amplitudes \cite{2}, and more recently, (d) nonclassical states of combined photon pairs also called $N00N$ states \cite{5,6}.

It is worth to note however, that both Schr\"odinger and Helmholtz equations are linear equations. However, in quantum optics, we have nonlinear Hamiltonians in the sense that powers of creation and annihilation operators (related to the quantized electromagnetic field) are considered. 

In this work, we show how to mimic a (quantum) Kerr medium by propagation of classical light through a quadratic GRIN medium. 
This is performed considering a second order correction to the paraxial propagation of light in the medium. In particular, the solutions for the non-paraxial wave equation produce nonlinear terms similar to the Hamiltonians proposed by Yurke and Stoler \cite{2} to generate Schr\"odinger cat states . 

We also predict the splitting of a Gaussian profile in two Gaussian functions. To the best of our knowledge,  this is  a new effect when one considers the propagation in quadratic GRIN media. 

We start with a quick survey of the solution of the Helmholtz equation for a quadratic GRIN medium in terms of eigenfunctions. The Helmholtz equation for a GRIN medium is
\begin{equation}
-\frac{\partial^2 E}{\partial z^2}=\left[    \frac{\partial^2}{\partial x^2}+\frac{\partial^2}{\partial y^2} + k^2 n^2(x,y) \right]  E,
\end{equation}
where $k$ is the wave number and $n(x,y)$ is the variable refraction index.
For a quadratic medium, the refraction index can be written as $n^2(x,y)=n_0^2\left( 1-g_x^2 x^2 -g_y^2 y^2\right) $, where $g_x$ and $g_y$ are the gradient indexes in the $x$ and in the $y$ directions, respectively. So, for a quadratic dependence in the index of refraction, the Helmholtz equation is expressed
\begin{equation}
-\frac{\partial^2 E}{\partial z^2}=\left[    \frac{\partial^2}{\partial x^2}+\frac{\partial^2}{\partial y^2} +\kappa^2-\eta_x^2 x^2-\eta_y^2 y^2 \right]  E,
\end{equation}
where we have defined $\kappa= n_0 k$, $\eta_x= n_0 g_x k$, and $\eta_y=n_0 g_y k$. Introducing the number operators $\hat{n}_x$ and $\hat{n}_y$, such that $\eta_\xi(\hat{n}_\xi+1/2)=\frac{1}{2}[(-i  \frac{d}{d\xi})^2+\eta_\xi^2 \xi ^2]\;(\xi=x,y)$, the previous equation can be cast as
\begin{equation}
\frac{\partial^2E}{\partial z^2}=-\left[ \kappa^2-\eta_x \left( 2 \hat{n}_x +1 \right) -\eta_y \left( 2 \hat{n}_y+1\right)\right] E,
\end{equation}
whose formal solution may be written as
\begin{equation}\label{Deaqui}
E(x,y,z)=e^{ -  i z \sqrt{\kappa^2-\eta_x \left( 2 \hat{n}_x +1 \right) -\eta_y \left( 2 \hat{n}_y+1\right)} }E(x,y,0).
\end{equation}
The boundary condition can be expanded in terms of the eigenfunctions
\begin{align}\label{eigenfun}
\varphi_n(\xi)= \left( \frac{\eta_\xi}{\pi}\right) ^{1/4}\frac{1}{\sqrt{2^n n!}}  \exp\left( -\frac{\eta_\xi}{2}\xi^2\right)   H_n(\sqrt{\eta_\xi} \, \xi),  \quad \xi=x,y, \quad n=0,1,2,...,
\end{align}
of the number operators $\hat{n}_x$ and $\hat{n}_y$ \cite{arfken}, where $H_n\left( \zeta \right) $ are the Hermite polynomials. Considering the explicit expansion
\begin{equation} \label{condin}
E(x,y,0)= \sum_{n=0}^{\infty} \sum_{m=0}^{\infty} c_{n,m} \varphi_n(x)\varphi_m(y),
\end{equation}
and applying the propagation operator (Eq. \eqref{Deaqui}), we arrive to the propagated field
\begin{align}\label{eq7}
E(x,y,z)= \sum_{n=0}^{\infty} \sum_{m=0}^{\infty} c_{n,m} \, \varphi_n(x) \, \varphi_m(y)  \exp\left[ -i z \sqrt{\kappa^2-\eta_x \left( 2 n +1 \right) -\eta_y \left( 2 m+1\right)}\right].
\end{align}

Now, to model a quantum Kerr medium, we use the Hamiltonian from \cite{2}
\begin{equation}
	H=\omega \, \hat{a}^{\dagger}\hat{a} +\mu (\hat{a}^{\dagger}\hat{a})^s,
\end{equation}
 to generate superpositions of coherent states of a quantized system. This Hamiltonian is nonlinear in the sense that creation and annihilation operators are related to the electromagnetic fields. Therefore, when having powers of these operators, we reach the nonlinear regime. As a particular case, when $s=2$, we are dealing with a quantum Kerr medium. In the above equation, $\omega$ is the field frequency and for $s=2$, $\mu$ is the Kerr parameter. Of course, what we want to do is to solve the Schr\"odinger equation (we set $\hbar=1$)
\begin{equation}
	i\frac{\partial |\psi(t)\rangle}{\partial t}=\hat{H}|\psi(t)\rangle,
\end{equation}
which is a linear equation. If we solve it for an initial coherent state \cite{Glauber}
\begin{equation}\label{estadocoherente}
|\alpha\rangle=e^{-\frac{|\alpha|^2}{2}}\sum_{m=0}^{\infty}\frac{\alpha^m}{\sqrt{m!}}|m\rangle,
\end{equation}
we obtain
\begin{align}\label{solu}
|\psi(t)\rangle=e^{-it[\omega \hat{a}^{\dagger}\hat{a} +\mu (\hat{a}^{\dagger}\hat{a})^s]}|\alpha\rangle =e^{-\frac{|\alpha|^2}{2}}\sum_{m=0}^{\infty}\frac{(\alpha e^{-i\omega t})^me^{-i\mu t m^s}}{\sqrt{m!}}|m\rangle.
\end{align}
Fo the particular value of time $t=\pi/\mu$, the second exponential in Eq. \eqref{solu} becomes 
$(-1)^{m^s}=(-1)^m$ and a coherent state is recovered. On the other hand, for the time $t=\pi/2\mu $, a superposition of coherent states is achieved \cite{2}, the so called Schr\"odinger cat states.

We can model the quantum Kerr medium by using Equation (\ref{Deaqui}), developing the square root as a Taylor series, and staying to second order in $1/\kappa^2$, obtaining the approximation
\begin{equation} \label{aproxi}
		E(x,y,z)\approx  
 		\exp\left\lbrace -i\tilde{\kappa}z\left[ 1-\frac{\eta_x\hat{n}_x+\eta_y\hat{n}_y}{\tilde{\kappa}^2}\right. \right. 
		 \left.\left.  -\frac{\left(\eta_x\hat{n}_x+\eta_y\hat{n}_y\right)^2}{2\tilde{\kappa}^4}\right]\right\rbrace E(x,y,0),
\end{equation}
with $\tilde{\kappa}^2=\kappa^2-(\eta_x+\eta_y)$. \\
Note that for s = 2, the exponential operator in Eq. \eqref{solu} presents both linear and squared number operators,  as occurs in Eq. \eqref{aproxi}. This formal similitude allows us to mimic classically the quantum interaction of light with matter that may be effectively modeled by a Kerr medium.\\
Without loss of generality and for the sake of simplicity, we consider the one-dimensional case, i.e. $\eta_y=0$; so, we have from \eqref{aproxi},
\begin{align} \label{aproxi2}
 E(x,z)=  \exp\left\lbrace -i\tilde{\kappa}z\left[ 1-\frac{\eta_x\hat{n}_x}{\tilde{\kappa}^2}-\frac{\left(\eta_x\hat{n}_x\right)^2}{2\tilde{\kappa}^4}\right]\right\rbrace E(x,0).
\end{align}
Now we assume that the boundary condition $E(x,0)$ has the form of the coherent state
\begin{equation} \label{coher}
\psi_\alpha\left( x \right)=e^{-\frac{|\alpha|^2}{2}}\sum_{m=0}^{\infty}\frac{\alpha^m}{\sqrt{m!}}\varphi_m(x),
\end{equation}
already defined in Eq. \eqref{estadocoherente}. Using the Hermite polynomials exponential generating function $\exp\left( 2xt-t^2\right)=\sum_{n=0}^{\infty} H_n(x) \frac{t^n}{n!} $ \cite{arfken, abramowitz}, valid for $x$ and $t$ complexes, it is very easy to see that $\psi_\alpha\left( x \right)$ is the Gaussian function \cite{Leonhardt},
\begin{align}\label{gauscoh}
	 \psi _{\alpha }(x)=\sqrt[4]{\frac{\eta _x}{\pi }} \exp \left\lbrace   -\frac{\eta _x}{2} \left[   x -\sqrt{\frac{2}{\eta_x}} \Re(\alpha) \right]^2 \right\rbrace 
	  \exp \left\lbrace i \sqrt{2 \eta_x} \Im \left( \alpha \right)  x   -  i \Re \left( \alpha \right) \Im \left( \alpha \right)        \right\rbrace,
\end{align}
where $\Re(\alpha) $ and $\Im(\alpha)$ are the real and imaginary part of $\alpha$, respectively.\\
Within the established conditions, it is now straightforward to obtain the propagated field as
\begin{align} \label{aproxi21}
 E(x,z)= e^{-i\tilde{\kappa} z} e^{-\frac{|\alpha|^2}{2}} \sum_{m=0}^{\infty}{\frac{\alpha^m }{\sqrt{m!}}} \exp\left[i z \left( \eta \, m + \chi \, m^2 \right)  \right] \varphi_m(x),
\end{align}
with $\eta=\eta_x/\tilde{\kappa}$  and $\chi=\eta^2_x/2\tilde{\kappa}^3$.\\
We note that in the paraxial case, where the quadratic term in the exponential of Eq. \eqref{aproxi21} is neglected,  $E(x,z)$ is periodic in $z$ with period $p_z=2\pi/\eta$. However, as it is shown below, such periodicity is also exhibited by the field $E(x,z)$, at least approximately, under non-paraxial conditions.

Comparing the propagation operator in Eq. \eqref{aproxi2} with the evolution operator in Eq. \eqref{solu}, it is noted that  $\mu t$ corresponds to $z \chi$. Therefore, the superposition of two coherent states in the quantum Kerr medium can be emulated in the quadratic GRIN device propagating the Gaussian field $\psi_\alpha$ to a distance $z_\mathrm{s}=\frac{\pi}{2\chi}$, and a revival of this field will occur at the  distance $z_\mathrm{r}=\frac{\pi}{\chi}$. Let us verify these predictions. \\
For $z_\mathrm{s}=\frac{\pi}{2\chi}$,  we have
\begin{align} \label{aproxi2b}
 E\left( x, \pi/2\chi  \right) =  e^{-i \tilde{\kappa} \pi/2\chi} e^{-\frac{|\alpha|^2}{2}} \sum_{m=0}^{\infty}{\frac{\alpha^m }{\sqrt{m!}}} e^{ i \pi \eta m/2\chi}\, e^{i \pi m^2/2}\varphi_m(x).
\end{align}
Next, we split the sum above into even and odd terms, we employ the identities  $( e^{i \pi/2})^{4m^2}=1$ and $(e^{i \pi/2})^{4m^2+4m+1}=i$, together with Eq. \eqref{coher},  and rearranging terms, we finally obtain
\begin{align} \label{aproxi4}
	E(x, \pi/2\chi  )= e^{-i\tilde{\kappa}\pi/2\chi} 	 \left[ \frac{1+i}{2} \psi_{\alpha e^{i\eta \pi/2\chi}}(x)+\frac{1-i}{2} \psi_{-\alpha e^{i\eta \pi/2\chi}}(x)\right] ,
\end{align}
i.e., the superposition of two Gaussian coherent states. \\
Following a procedure similar to the previous one, we get that for $z_\mathrm{r}=\frac{\pi}{\chi}$ the field \eqref{aproxi21} can be expressed as
\begin{equation}\label{revival}
E(x,  \pi/\chi )= e^{-i\tilde{k}\pi/\chi} \; \psi_{-\alpha e^{i\eta \pi/\chi}}(x),
\end{equation}
that is a coherent state multiplied by a phase.\\
\begin{figure}
 \centering
\begin{subfigure}{.45\textwidth}
  \centering
  \includegraphics[width=.8\linewidth, height=0.2\textheight]{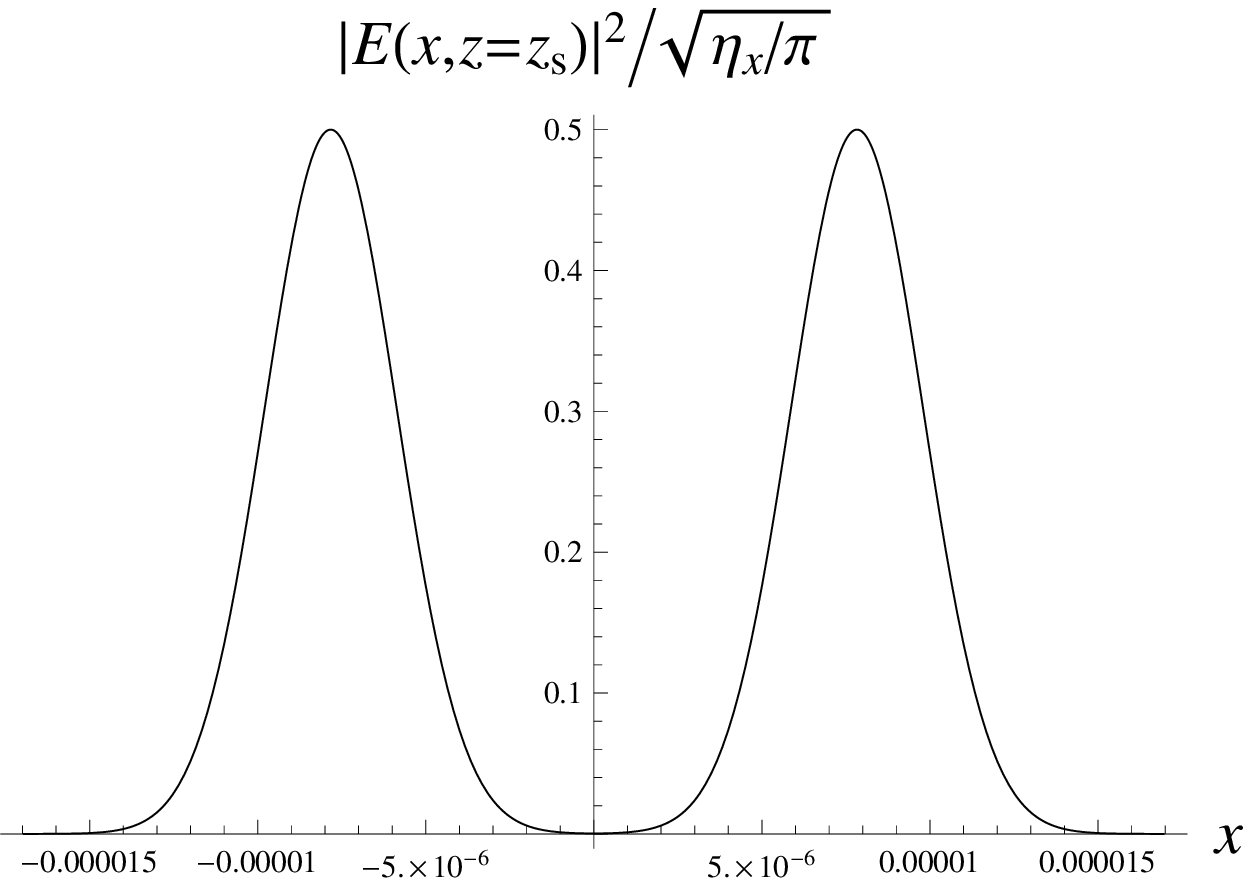}
  \caption{}
  \label{fig2d1a}
\end{subfigure}
\begin{subfigure}{.45\textwidth}
	\centering
  \includegraphics[width=.8\linewidth, height=0.2\textheight]{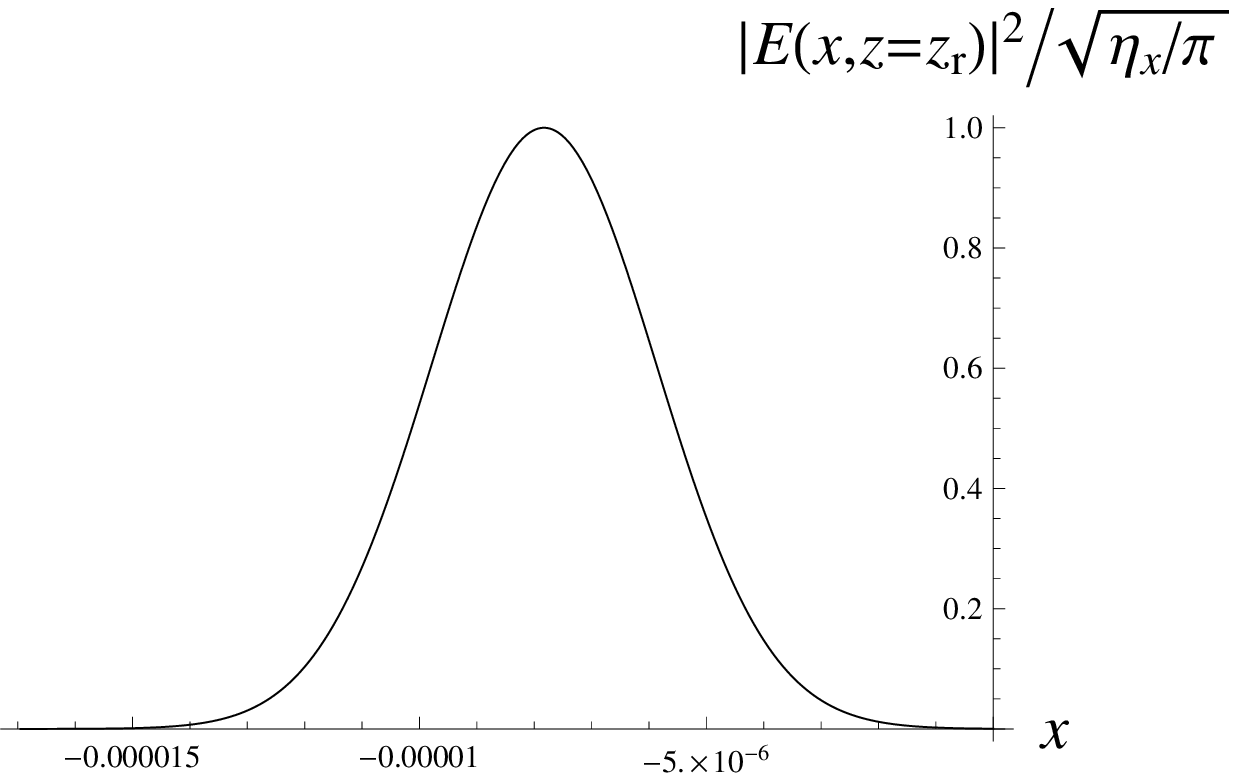}
  \caption{}
  \label{fig2d1b}
\end{subfigure}
\caption{Normalized intensity of (a) the splitting and (b) the revival fields for $n_0=1.5$, $g_x=10 \; \mathrm{mm}^{-1}$, $k=8.7 \times 10^6 \; \mathrm{m}^{-1}$ and $\alpha=2$.}
\label{fig2d1}
\end{figure}
To illustrate these results, we consider a coherent state with $\alpha=2$ and $k=8.7 \times 10^6 \; \mathrm{m}^{-1}$, propagating in a GRIN medium with parameters $n_0=1.5$ and $g_x=10 \; \mathrm{mm}^{-1}$. In this case, the splitting and the revival fields appear at distances $z_\mathrm{s}=40.95 \; \mathrm{cm}$ and $z_\mathrm{r}=81.90 \; \mathrm{cm}$, respectively. The normalized intensity profiles of the split and revival fields are depicted in Fig. \ref{fig2d1}.\\
The intensity of the field $E(x,z)$, Eq. \eqref{aproxi21}, obtained for the previously considered parameters,  is displayed in Fig. \ref{fig3d}, for three different intervals of $z$, whose length is equal to the period $p_z$. For the considered parameters this period is $p_z=0.628 \; \mathrm{mm}$. The first interval, starting at $z=0$, present a field that is also predicted by the paraxial approximation. The second and third intervals, centered respectively at $z=z_\mathrm{s}$ and $z=z_\mathrm{r}$, correspond to the split and revival fields, that have been predicted in Eqs. \eqref{aproxi4} and \eqref{revival}.  The transverse field in the revival zone coincides with the field at $z=0$ only at the plane $z=z_\mathrm{r}$. It is noted that the paraxial approximation predicts that the field evolution in the splitting and revival intervals of Fig. \ref{fig3d}, are identical to that in the first interval.\\
\begin{figure}
 \centering
\begin{subfigure}{.32\textwidth}
  \centering
  \includegraphics[width=.9\linewidth, height=0.16\textheight]{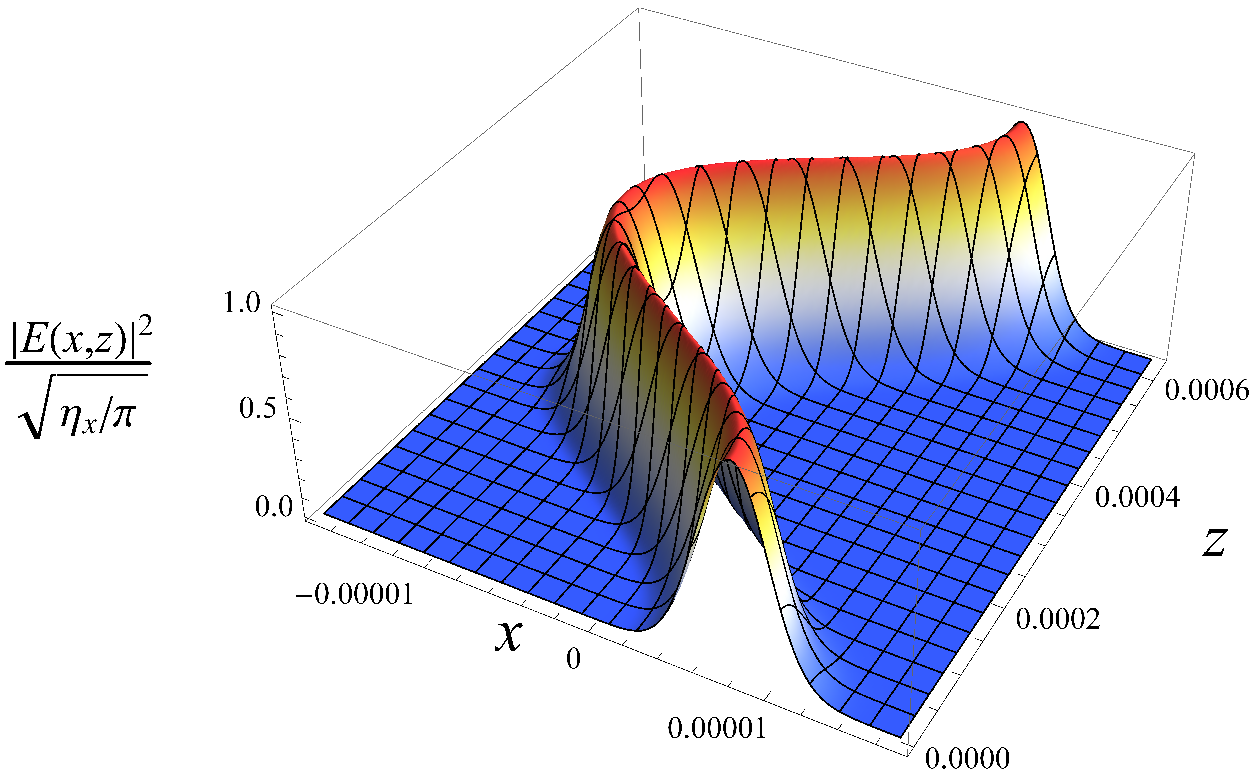}
  \caption{}
  \label{fig3da}
\end{subfigure}
\begin{subfigure}{.32\textwidth}
	\centering
  \includegraphics[width=.9\linewidth, height=0.16\textheight]{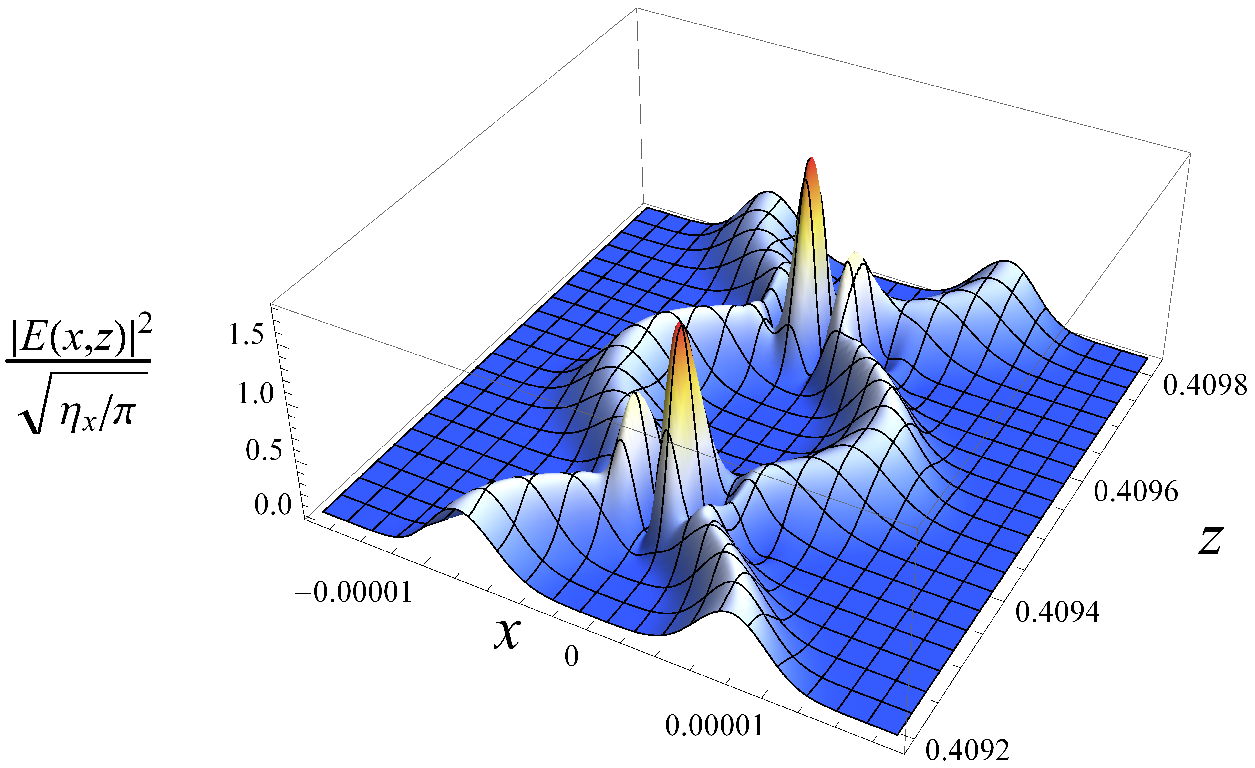}
  \caption{}
  \label{fig3db}
\end{subfigure}
\begin{subfigure}{.32\textwidth}
	\centering
  \includegraphics[width=.9\linewidth, height=0.16\textheight]{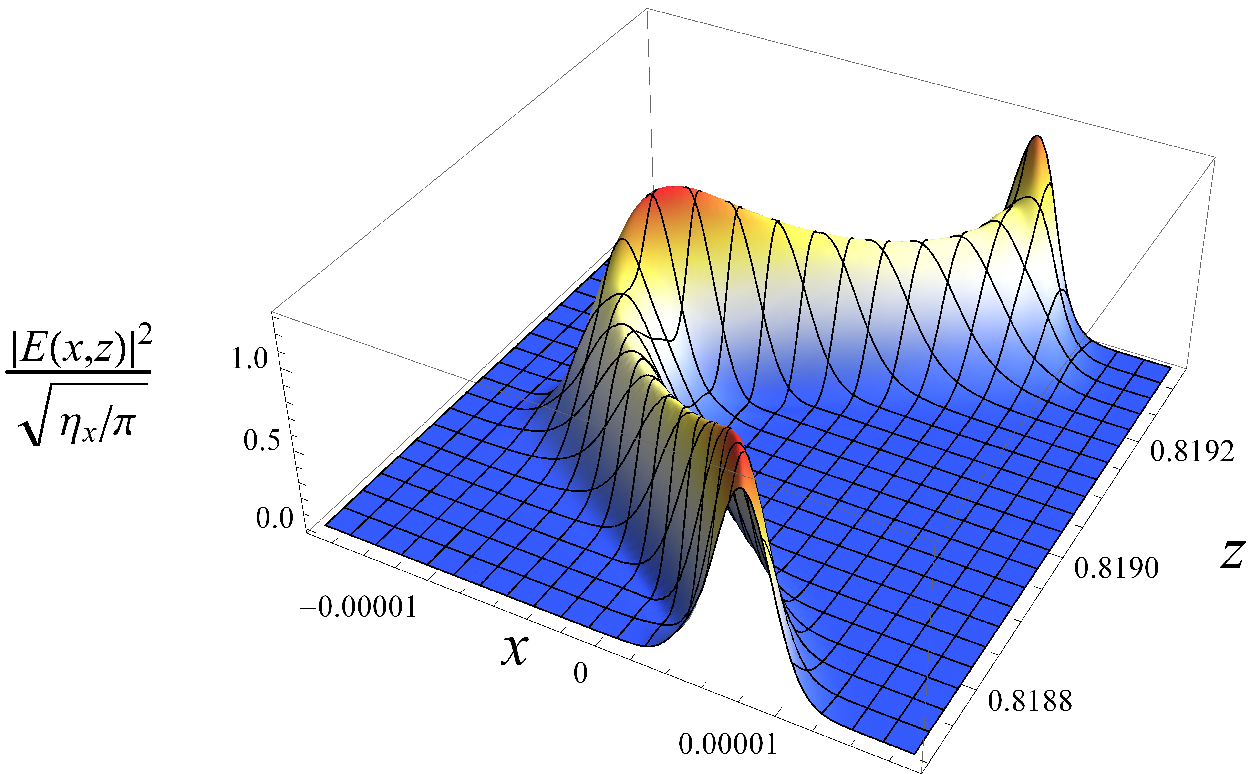}
  \caption{}
  \label{fig3dc}
\end{subfigure}
\caption{Intensity profiles at different intervals of $z$, of length $p_z$, obtained for  $n_0=1.5$, $g_x=10 \; \mathrm{mm}^{-1}$, $k=8.7 \times 10^6 \; \mathrm{m}^{-1}$ and $\alpha=2$. The intervals are centered at (a) $z=p_z/2$, (b) $z=z_\mathrm{s}$, and (c) $z=z_\mathrm{r}$.}
\label{fig3d}
\end{figure}
The condition for the validity of the approximation in  Eq. \eqref{aproxi2} is $\frac{2 g_x }{k n_0-g_x}<<1$. The parameters chosen in the previous numerical illustration allow the fulfillment of such approximation. Indeed, the computation of the split and revival fields, using the one dimensional version of the exact formula (Eq. \eqref{eq7}), generates results almost identical to the ones shown in Figs. \ref{fig2d1} and \ref{fig3d}.
The chosen value of $k$, in the visible spectral domain, minimizes the interference terms that arise squaring \eqref{aproxi4}.\\
When the condition $\frac{2 g_x }{k n_0-g_x}<<1$ is not fulfilled, the split and revival fields do not appear. An example of this situation occurs for an initial coherent state with $\alpha=2$ and $k=10^5 \; \mathrm{m}^{-1}$ propagating in a medium with the parameters $n_0=1.5$ and $g_x=10 \; \mathrm{mm}^{-1}$. The normalized intensity profile at the splitting distance ($z_\mathrm{s}=4.25 \; \mathrm{mm}$ in this case), shown in Fig. \ref{fig2d2}, does not correspond to the predicted field in Eq. \eqref{aproxi4}.
\begin{figure}[h!]
\centering
\includegraphics[width=0.5\linewidth, height=0.25\textheight]{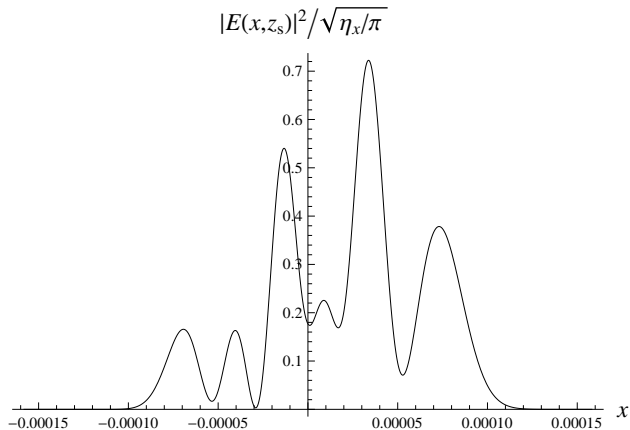}
\caption{Normalized intensity at $z=z_\mathrm{s}$ for  $n_0=1.5$, $k=10^5\; \mathrm{m}^{-1}$, $g_x=10 \; \mathrm{mm}^{-1}$ and $\alpha=2$.}
\label{fig2d2}
\end{figure}

We conclude that by propagating light though a GRIN medium with transverse quadratic variation, we have been able to model the quantum interaction of a quantized field with a nonlinear (Kerr) medium. We have achieved this by writing the Helmholtz equation in operator terms, and approximating the solution up to second order. The ``Hamiltonian" that we produced was exactly the Kerr medium Hamiltonian used in quantum optics, that generates superpositions of coherent states of the quantized field. 

We have also shown the splitting of a Gaussian field, with the appropriate width and position, during its  propagation in a quadratic  GRIN media.
According to data in different papers \cite{AO1,AO2,OL1}, the values of the gradient parameter $g$, for conventional GRIN devices operating in the visible domain, are in the range from $0.1$ to $10.0  \: \mathrm{mm}^{-1}$.  
In the example discussed above, where we employed the gradient parameter $g=10 \; \mathrm{mm}^{-1}$ and $k=8.7 \times 10^{6} \; \mathrm{m}^{-1}$  ($\lambda=722 \: \mathrm{nm}$) the splitting distance is 40.95 cm. 
Therefore, the experimental verification of the splitting effect could be implemented under certain limit conditions employing present GRIN technology. 
The splitting distance can be reduced increasing the wavelength without affecting the validity of Eq. \eqref{aproxi2}. For instance, employing $\alpha=1$ and $\lambda=1.57 \; \mathrm{\mu m}$ for the initial coherent state, and $n_0 = 1.5$ and $g_x=10 \; \mathrm{mm}^{-1}$ for the parameters of the medium, the splitting distance is 18.8 cm.


\begin{thebibliography}{XX}
 \bibitem{gomez} C. Gomez-Reino, M.V. Perez, and C. Bao; \textit{Gradient-Index Optics}, Springer-Verlag, 2002.
 \bibitem{ozaktas} H.M. Ozaktas and D. Mendlovic; "Fractional Fourier transforms and their optical implementation. II", \textit{J. Opt. Soc. Am.} A 10, 2522-2531 (1993).
 \bibitem{ojeda} J. Ojeda-Casta\~{n}eda and P. Szwaykowski; \textit{Novel Modes in 2-power GRIN}, SPIE Vol. 1500, Innovative Optics and Phase Conjugate Optics, 246-251 (1991).
 \bibitem{Arr} H.M. Moya-Cessa, M. Fern\'andez Guasti, V.M. Arrizon and S. Ch\'avez-Cerda; "Optical realization of quantum-mechanical invariants", {\it Opt. Lett.} {\bf 34}, 1459 (2009).
 \bibitem{Chuma} S.M. Chumakov and K.B. Wolf; "Supersymmetry in Helmholtz optics", \textit{Phys. Lett.} A {\bf 193}, 51 (1994).
 \bibitem{Dimi} M.A. Miri, M. Heinrich, R. El-Ganainy, and D.N. Christodoulides; "Supersymmetric optical structures", \textit{Phys. Rev. Lett.} 110, 233902 (2013).
 \bibitem{David} A. Z\'u\~niga-Segundo, B.M. Rodr\'{\i}guez-Lara, D.J. Fern\'andez and H.M. Moya-Cessa; "Jacobi photonic lattices and their SUSY partners", \textit{Opt. Express} {\bf 22}, 987 (2014).
 \bibitem{loudon} R. Loudon and P.L. Knight; "Squeezed light",  \textit{Special Issue of  J. of Mod. Opt.} {\bf 34}, 709 (1987).
 \bibitem{4} J. Krause, M.O. Scully, T. Walther, and H. Walther; "Preparation of a pure number state and measurement of the photon statistics in a high-Q micromaser", \textit{Phys. Rev.} A {\bf 39}, 1915 (1989).
 \bibitem{1} E. Schr\"odinger; “Die gegenwärtige Situation in der Quantenmechanik”, \textit{Die Naturwissenschaften} 23, 807-812, 823-828, 844-849 (1935).
 \bibitem{2} B. Yurke and D. Stoler; "Generating quantum mechanical superpositions of macroscopically distinguishable states via amplitude dispersion", \textit{Phys. Rev. Lett.} {\bf 57}, 13 (1986).
 \bibitem{5} I. Afek, O.Ambar, and Y. Silberberg; "High-NOON States by Mixing Quantum and Classical Light", \textit{Science} {\bf 328}, 879 (2010).
 \bibitem{6} J.P. Dowling; "Quantum optical metrology – the lowdown on high-N00N states", \textit{Contemp. Phys.} 49, 125 (2008).
 \bibitem{arfken} G.B. Arfken, H.J. Weber and F.E. Harris. \textit{Mathematical Methods for Physicist}. Elsevier 2013.
 \bibitem{Glauber} R.J. Glauber; "Coherent and Incoherent States of the Radiation Field ",  Phys. Rev. A{\bf 131}, 2766 (1963).
 \bibitem{abramowitz} M. Abramowitz and I.A. Stegun. \textit{Handbook of Mathematical Functions}. NBS 1972. 
 \bibitem{Leonhardt} U. Leonhardt, {\it Measuring the Quantum State of Light}, Cambridge University Press, 1997.
 \bibitem{AO1} K. Iga; "Theory of gradient-index imaging", Appl. Opt. 19, 1039-1043 (1980).
 \bibitem{AO2} B. Messerschmidt, T. Possner, and R. Goering; "Colorless gradient-index cylindrical lenses with high numerical apertures produced by silver-ion exchange", Appl. Opt. 34, 7825-7830 (1995).
 \bibitem{OL1} S.H. Song, S. Park, C.H. Oh, and P.S. Kim; "Gradient-index planar optics for optical interconnections", Opt. Lett. 23, 1025-1027 (1998).
 \end{thebibliography}
\end{document}